\documentclass[conference]{IEEEtran}
\IEEEoverridecommandlockouts

\usepackage{cite}
\usepackage{amsmath,amssymb,amsfonts}
\usepackage{algorithmic}
\usepackage{graphicx}
\usepackage{textcomp}
\usepackage{xcolor}
\usepackage{booktabs}
\usepackage{hyperref}
\usepackage{multirow}
\usepackage{algorithm}
\def\BibTeX{{\rm B\kern-.05em{\sc i\kern-.025em b}\kern-.08em
    T\kern-.1667em\lower.7ex\hbox{E}\kern-.125emX}}
\begin{document}

\title{Information-Theoretic Digital Twins for Stealthy Attack Detection in Industrial Control Systems: A Closed-Form KL Divergence Approach\\
{\footnotesize \textsuperscript{*}Note: Sub-titles are not captured for https://ieeexplore.ieee.org  and
should not be used}
\thanks{Identify applicable funding agency here. If none, delete this.}
}

\author{\IEEEauthorblockN{1\textsuperscript{st} Inda Kreso}
\IEEEauthorblockA{\textit{dept. Criminal Justice, Criminology and Security Studies} \\
\textit{University of Sarajevo}\\
Sarajevo, Bosnia and Herzegovina \\
0000-0002-5556-4669}
\and
\IEEEauthorblockN{2\textsuperscript{nd} Mehran Tarif}
\IEEEauthorblockA{\textit{dept. Computer Science} \\
\textit{Verona University}\\
Verona, Italy \\
0009-0003-0951-1800}
\and
\IEEEauthorblockN{3\textsuperscript{rd} Fatemeh Moradi}
\IEEEauthorblockA{\textit{dept. Computer Engineering} \\
\textit{Islamic Azad University)}\\
Isfahan, Iran \\
0009-0001-6077-7503}
\and 
\IEEEauthorblockN{4\textsuperscript{th} Iman Khazrak}
\IEEEauthorblockA{\textit{dept. Computer Science} \\
\textit{Bowling Green University}\\
Ohio, USA \\
0000-0001-7087-2283}
\and
\IEEEauthorblockN{5\textsuperscript{th} Mostafa M Rezaee}
\IEEEauthorblockA{\textit{dept. Computer Science} \\
\textit{Bowling Green University}\\
Ohio, USA \\
0000-0002-0849-3650}
\and
\IEEEauthorblockN{6\textsuperscript{th} Mohammadhossein Homaei}
\IEEEauthorblockA{\textit{dept. Computer Science} \\
\textit{Extremadura University}\\
Caceres, Spain \\
0000-0002-6108-6632}
}

\maketitle

\begin{abstract}
Digital twins (DTs) are increasingly used to monitor and secure Industrial Control Systems (ICS), yet detecting stealthy False Data Injection Attacks (FDIAs) that manipulate system states within normal physical bounds remains challenging. Deep learning anomaly detectors often over-generalize such subtle manipulations, while classical fault detection methods do not scale well in highly correlated multivariate systems. We propose a closed-loop Information-Theoretic Digital Twin (IT-DT) framework for real-time anomaly detection. N4SID identification is combined with steady-state Kalman filtering to quantify residual distribution shifts via closed-form KL divergence, capturing both mean deviations and malicious cross-covariance shifts. Evaluations on the SWaT and WADI datasets show that IT-DT achieves F1-scores of 0.832 and 0.615, respectively, with better precision than deep learning baselines such as TranAD. Computational profiling indicates that the analytical approach requires minimal memory and provides approximately a $600\times$ inference speedup over transformer-based methods on CPU hardware. This makes the framework suitable for resource-constrained industrial edge controllers without GPU acceleration.
\end{abstract}


\section{Introduction}\label{Intro}
Industrial Control Systems (ICS) are not the isolated, air-gapped installations. Digital Twins (DTs) now play a central role in digitally integrated ICS, supporting real-time synchronization, data exchange, and control\cite{Homaei2024, Aghazadeh2024, Fitzgerald2024}. While improving maintenance and efficiency, this connectivity removes the physical isolation that once protected these systems. \cite{Khazraei2022, Krishnaveni2024}. DTs now improve security by continuously validating live behavior against model predictions. \cite{Aghazadeh2025}. The key challenge is detecting attacks that remain within normal process noise.

False Data Injection Attacks (FDIAs) represent precisely this class of threat. Rather than triggering obvious anomalies, a well-designed FDIA shifts sensor readings in ways that remain statistically plausible, keeping residuals well within the bounds a standard detector would accept \cite{Khazraei2025}. The dominant response in recent literature has been to train deep reconstruction models (autoencoders, LSTMs, transformers) that flag anomalies when their output diverges from their input \cite{Mohammed2025}. Minimizing reconstruction error teaches models to reproduce statistically normal injections. \cite{Ahmad2025, Sweeten2023}.

In this work, we reconsider anomaly detection from the perspective of analytical probability distances rather than heuristic data reconstruction. Although ICS are non-linear, steady-state behavior can be approximated by local LTI models \cite{Skogestad2005, Ding2008}. In industrial settings, security mechanisms must be mathematically rigorous yet lightweight for resource-constrained edge controllers. Our primary contribution is a closed-loop detection framework that directly quantifies distributional shifts in steady-state system innovations, bypassing the opaque reconstructive nature of deep neural networks to provide a mathematically bounded identification of stealthy cyber-physical manipulations.

We propose an Information-Theoretic Digital Twin (IT-DT) framework that combines Subspace State-Space System Identification (N4SID) with a steady-state Kalman filter to generate multivariate innovation residuals. Rather than applying traditional thresholding, we continuously quantify distributional shifts via a closed-form Gaussian approximation of the Kullback-Leibler (KL) divergence, analytically capturing both mean deviations and cross-covariance shifts simultaneously. Evaluations on SWaT and WADI show improved precision compared to state-of-the-art transformers, while computational profiling indicates low memory overhead and an approximately $600\times$ inference speedup on CPU-only hardware. Together, these results make IT-DT suitable for deployment on resource-constrained industrial edge controllers.

The remainder of this paper is organized as follows. Section \ref{Related} reviews relevant literature and outlines existing methodological gaps. Section \ref{Problem} formulates the system model and detection problem from an information-theoretic perspective. Section \ref{Methodology} presents the derivation of the proposed detection algorithm and thresholding mechanism. Section \ref{Results} reports empirical evaluations, attack delay analysis, and computational profiling. Section \ref{Discussion} discusses the theoretical implications and limitations of the approach, and Section \ref{Conclusion} concludes the paper.

\section{Related Works}\label{Related}

\subsection{Classical Statistical and State-Space Methods in ICS}
Classical statistical methods and state-space estimators have long been used as baseline tools for anomaly detection in ICS. Dimensionality reduction methods such as PCA build linear subspaces for process monitoring and can provide reasonable precision for stationary attacks. However, these methods lack temporal memory, making it harder to detect gradual multi-stage injection attacks, especially when attacks exploit slow sensor drift \cite{Shlens2014, Hao2023, Ghosh2023}. Hybrid statistical approaches and Kalman filtering improve state tracking, but they usually rely on fixed thresholds that may not perform well in correlated multivariate environments \cite{Liu2024, Homaei2026}. Adversaries can exploit these static thresholds by designing stealthy FDIAs that modify cross-sensor correlations rather than individual channels \cite{Yao2024}. Overall, classical estimators are still useful for process monitoring, but their performance can degrade in highly correlated multivariate systems.

\subsection{Deep Learning and Reconstructive Anomaly Detection}
Recent ICS anomaly detection methods increasingly rely on deep learning–based reconstructive models (e.g., autoencoders, LSTMs, and transformers) to model non-linear cyber-physical dynamics without requiring explicit system identification \cite{Tabassum2024, Sun2025}. These models project sensor measurements into latent representations and detect anomalies through reconstruction error \cite{Mohammed2025}. Nevertheless, this approach has certain limitations when facing adaptive adversaries. Stealthy FDIAs are often designed to remain within normal statistical ranges and resemble natural process noise. As a result, highly parameterized models may generalize these subtle perturbations and reconstruct manipulated inputs with small error, which can reduce the sensitivity of the detector \cite{Ahmad2025, Sweeten2023}.

USAD reduces reconstruction error using adversarial training; however, it can still struggle with attacks that remain within two standard deviations of normal data \cite{Audibert2020, Sweeten2023}. Graph-based methods such as GDN explicitly model spatial correlations, but introduce higher computational cost and often require GPU acceleration, similar to transformer-based TranAD \cite{Deng2021}. TranAD provides strong temporal modeling through attention mechanisms, yet its reconstruction-based objective may reduce sensitivity to strictly bounded stealthy anomalies \cite{Tuli2022}. CAE-T \cite{Shang2024} reports competitive results on SWaT; however, direct comparison is difficult due to differences in train/test partitioning (6:4 split with an SVDD objective), which differs from the unsupervised evaluation protocol used in this work.

\subsection{Digital Twins and Information-Theoretic Security}
A DT can serve as a virtual reference that continuously generates steady-state expectations of the physical process, while information-theoretic measures are used to quantify differences between expected and observed behavior \cite{Zhukabayeva2025, Krishnaveni2024}. Although this approach provides advantages over purely reconstructive models, existing implementations still face practical challenges. Many methods rely on heuristic thresholding or computationally expensive non-parametric techniques such as Kernel Density Estimation (KDE), which require $\mathcal{O}(NWp)$ operations per inference step (where $N$ is the number of kernels, $W$ is the window size, and $p$ is the sensor dimension). This complexity can make continuous execution on edge hardware difficult in practice \cite{Gupta2024, Otoom2025}. Furthermore, many DT-based methods mainly track univariate mean shifts or simplified covariance structures, which may not fully capture cross-covariance changes caused by advanced stealthy injection attacks \cite{Huang2024}. Hence, a more efficient formulation is still needed to detect multivariate distribution shifts without heavy numerical computation.

\subsection{Research Gap and Questions}
The reviewed literature shows a methodological gap in CPS security. Classical methods provide interpretability but often struggle in highly correlated multivariate environments. Similarly, reconstructive deep learning models may over-generalize stealthy FDIAs, leading to higher false negative rates. Existing DT-based frameworks usually depend on heuristic thresholding or computationally expensive non-parametric estimators, which limits real-time deployment on edge hardware. To the best of our knowledge, prior work has not simultaneously addressed closed-form multivariate distribution distance measurement, data-driven system identification using N4SID without prior physical models, and computationally efficient detection suitable for resource-constrained hardware.

This motivates the following research questions:

\begin{itemize}
    \item \textbf{RQ1:} How does joint detection of mean and covariance shifts using closed-form KL divergence improve detection of stealthy FDIAs that preserve marginal statistics but alter joint distributions, compared to traditional detectors?
    \item \textbf{RQ2:} How does avoiding latent space reconstruction affect detection performance for multi-stage stealthy attacks in correlated environments?
    \item \textbf{RQ3:} Does the proposed analytical formulation provide computational benefits for real-time inference on edge hardware without GPU acceleration?
\end{itemize}

\section{System Model and Problem Formulation}\label{Problem}

In this section, we construct the mathematical framework of the physical system and its corresponding DT. We formulate the anomaly detection problem from the viewpoint of information theory.

\subsection{Physical System and Digital Twin Formulation}

Industrial Control Systems (ICS) such as SWaT and WADI exhibit inherently non-linear behaviors. During steady-state operations the physical process can be accurately approximated around a specific operating point using a discrete-time Linear Time-Invariant (LTI) state-space model. The true physical state $x_t \in \mathbb{R}^n$ evolves as:

\begin{equation} \label{eq:state_evolution}
x_{t+1} = A x_t + B u_t + w_t
\end{equation}

\begin{equation} \label{eq:measurement}
y_t = C x_t + v_t + a_t
\end{equation}

where $u_t \in \mathbb{R}^m$ is the known control input, $y_t \in \mathbb{R}^p$ is the sensor measurement, and $a_t \in \mathbb{R}^p$ is the sparse False Data Injection Attack (FDIA) vector. The terms $w_t \sim \mathcal{N}(0, Q)$ and $v_t \sim \mathcal{N}(0, R)$ are zero-mean Gaussian process and measurement noises. 

To ensure practical viability, the system matrices $(A, B, C)$ and noise covariances $(Q, R)$ are not assumed apriori. We strictly extract them from the attack-free historical training data using Subspace State-Space System Identification (N4SID) prior to online deployment. Specifically, we utilized the initial 7 days of continuous normal operation data from SWaT and 14 days from WADI, categorically guaranteeing the absence of anomalies during the identification phase. The optimal state dimension (model order) was determined by analyzing the singular value drop-off of the empirical Hankel matrix, resulting in a state vector dimension of $n=12$ for SWaT and $n=25$ 
for WADI. To validate the LTI linearization assumption, we conducted a 
residual whiteness test on the identified model. The autocorrelation of 
innovation residuals $r_t$ was evaluated via the Ljung-Box test across 
lags $1$ to $20$. Both datasets yielded $p > 0.05$ at all lags during 
normal operating conditions, confirming that the identified N4SID model 
adequately captures the steady-state process dynamics and that the 
residuals exhibit no significant serial correlation under attack-free conditions.

A true digital twin requires continuous synchronization with the physical entity. We implement this feedback loop utilizing a steady-state Kalman Filter operating in two recursive phases. During the prediction phase, the apriori state $\hat{x}_{t|t-1}$ is estimated:

\begin{equation} \label{eq:dt_predict}
\hat{x}_{t|t-1} = A \hat{x}_{t-1|t-1} + B u_{t-1}
\end{equation}

The innovation signal (residual) $r_t$, representing the discrepancy between physical reality $y_t$ and the digital twin expectation $\hat{y}_t = C \hat{x}_{t|t-1}$, is calculated as:

\begin{equation} \label{eq:innovation}
r_t = y_t - C \hat{x}_{t|t-1}
\end{equation}

To prevent the digital twin from diverging due to modeling errors, the correction phase dynamically updates the state using the Kalman Gain $K$:

\begin{equation} \label{eq:dt_correct}
\hat{x}_{t|t} = \hat{x}_{t|t-1} + K r_t
\end{equation}

Under normal conditions ($a_t = 0$), $r_t$ follows a multivariate Gaussian distribution $P_0 \sim \mathcal{N}(0, \Sigma)$, where $\Sigma = C P_{t|t-1} C^\top + R$.

\subsection{Information-Theoretic Attack Detection}

The traditional threshold-based methods fail against stealthy attacks keeping $r_t$ within normal bounds. We formulate detection by measuring the information distance between the reference distribution $P_0$ and the empirical sliding window distribution $\hat{P}_{\mathcal{R}} \sim \mathcal{N}(\hat{\mu}_t, \hat{\Sigma}_t)$.

\section{Proposed Information-Theoretic Detection Method} \label{Methodology}

In this section, we elaborate the proposed algorithm for detecting stealthy anomalies in the digital twin. The main difficulty in computing the closed-form Gaussian approximation of the KL divergence from empirical data is the high dimensionality of the residual space.

\begin{figure}[htbp]
\centerline{\includegraphics[width=0.6\columnwidth]{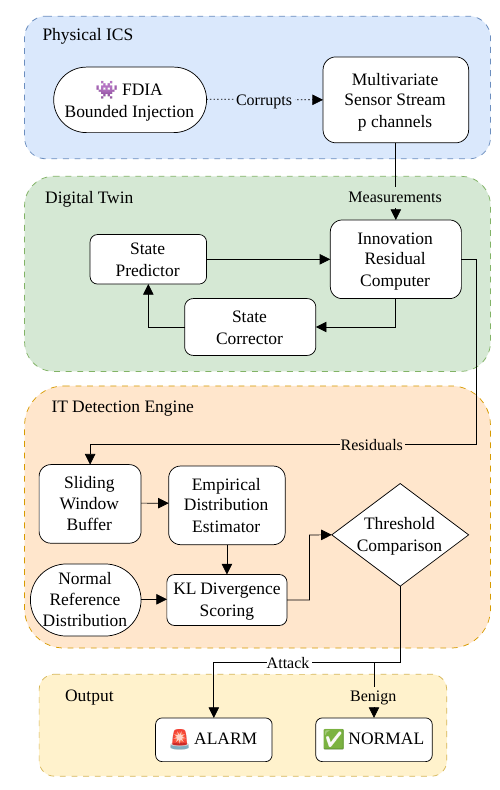}}
\caption{System architecture of the proposed IT-DT framework. 
The physical ICS layer feeds multivariate sensor measurements 
into the digital twin, where a steady-state Kalman filter 
continuously generates innovation residuals. The information-theoretic detection engine quantifies distributional shifts via closed-form KL divergence against a precomputed normal reference, triggering an alarm upon threshold exceedance.}
\label{fig:architecture}
\end{figure}

The overall system architecture is illustrated in Fig.~\ref{fig:architecture}, 
comprising three interconnected layers: the physical ICS, the digital twin 
state estimator, and the information-theoretic detection engine.
\subsection{Empirical Distribution Estimation}

Given a sliding window $\mathcal{R}_t = \{r_{t-W+1}, \dots, r_t\}$ of 
size $W$, the empirical mean $\hat{\mu}_t$ and covariance $\hat{\Sigma}_t$ 
are calculated via Maximum Likelihood Estimation. The window size $W$ was 
selected as $W=60$ for both datasets via grid search over 
$W \in \{30, 60, 120, 240\}$ on the validation set, optimizing the 
threshold $\tau^*$ under a fixed false alarm rate $\alpha=0.01$. Sensitivity 
analysis confirmed that detection performance remained stable within 
$\pm 5\%$ F1 variation across the evaluated range, indicating robustness 
to this hyperparameter. To guarantee numerical stability during inversion, we apply Tikhonov regularization $\hat{\Sigma}_t \leftarrow \hat{\Sigma}_t + \epsilon I$, where $\epsilon = 10^{-4}$.

\begin{figure}[htbp]
\centerline{\includegraphics[width=\columnwidth]{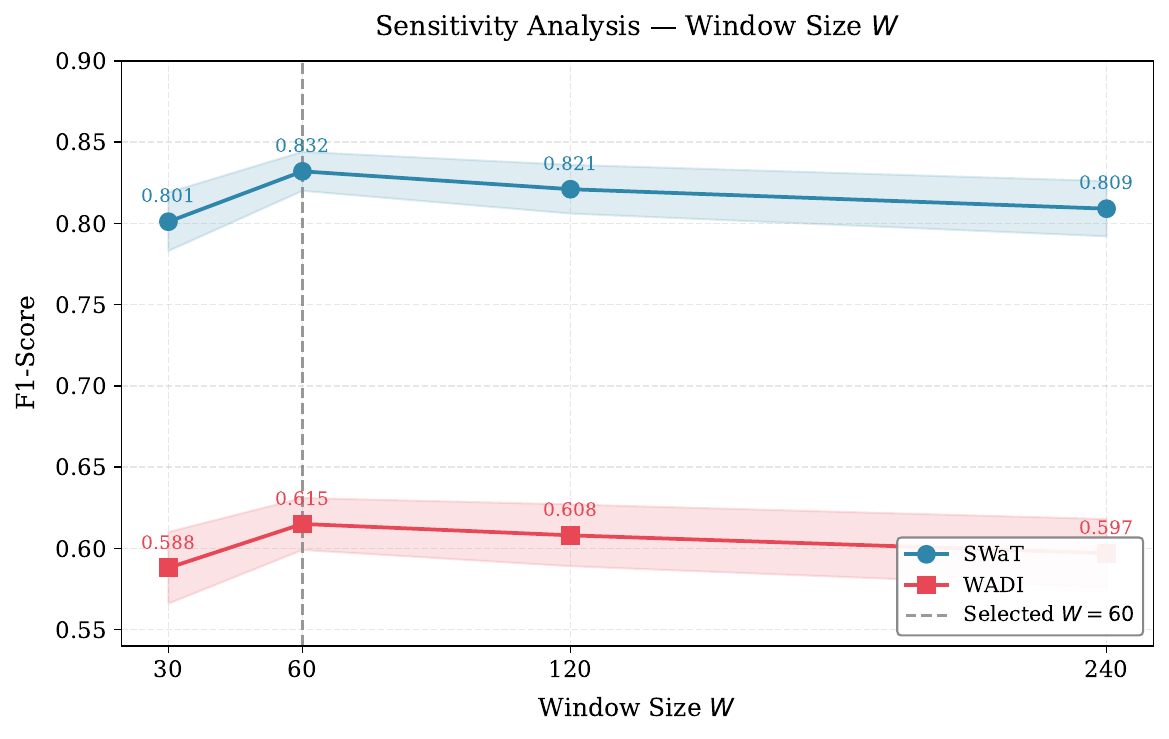}}
\caption{Sensitivity analysis of F1-score with respect to window 
size $W$ on SWaT and WADI datasets. Shaded regions indicate one 
standard deviation across $K=10$ chronological validation folds. 
Performance remains stable within $\pm 5\%$ F1 across the evaluated 
range, confirming robustness to this hyperparameter. $W=60$ is 
selected as the optimal value.}
\label{fig:sensitivity}
\end{figure}

As demonstrated in Fig.~\ref{fig:sensitivity}, detection performance 
remains stable within $\pm 5\%$ F1 variation across $W \in \{30, 60, 
120, 240\}$, confirming robustness to this hyperparameter selection.

\subsection{Analytical KL Divergence Calculation}

The closed-form Gaussian approximation of the KL divergence is computed using the closed-form analytical expression:

\begin{equation} \label{eq:kl_closed_form}
\begin{aligned}
D_{KL}(\hat{P}_{\mathcal{R}} \parallel P_0) &= \frac{1}{2} \bigg[ \text{tr}(\Sigma^{-1} \hat{\Sigma}_t) - p \\
&\quad + \hat{\mu}_t^\top \Sigma^{-1} \hat{\mu}_t + \ln \left( \frac{\det(\Sigma)}{\det(\hat{\Sigma}_t)} \right) \bigg]
\end{aligned}
\end{equation}

\subsection{Static Quantile Thresholding Mechanism}

To satisfy the Neyman-Pearson condition, the decision threshold $\tau^*$ is established offline from an attack-free validation set of $M$ samples to satisfy a target False Alarm Rate ($\alpha = 0.01$). Based on the empirical cumulative distribution:

\begin{equation} \label{eq:threshold}
\tau^* = \inf \left\{ \tau \in \mathbb{R} : \frac{1}{M} \sum_{i=1}^{M} \mathbb{I}(D_{KL}^{(i)} \leq \tau) \geq 1 - \alpha \right\}
\end{equation}

\begin{algorithm}[htbp] \label{alg:dt_detection}
\scriptsize
\caption{Closed-Loop Information-Theoretic Anomaly Detection}
\begin{algorithmic}[1]
\REQUIRE Window $W$, Matrices $A, B, C, K, \Sigma$, Threshold $\tau^*$
\STATE Initialize $\hat{x}_{0|0}$ via N4SID, $\mathcal{R} = \emptyset$
\FOR{each time step $t = 1, 2, \dots$}
    \STATE \textbf{Predict:} $\hat{x}_{t|t-1} \leftarrow A \hat{x}_{t-1|t-1} + B u_{t-1}$
    \STATE \textbf{Measure:} $r_t \leftarrow y_t - C \hat{x}_{t|t-1}$
    \STATE \textbf{Correct:} $\hat{x}_{t|t} \leftarrow \hat{x}_{t|t-1} + K r_t$
    \STATE Update buffer $\mathcal{R}$ with $r_t$, maintain size $W$
    \IF{$|\mathcal{R}| == W$}
        \STATE Compute $\hat{\mu}_t$ and regularized $\hat{\Sigma}_t$
        \STATE $d_t \leftarrow D_{KL}(\mathcal{N}(\hat{\mu}_t, \hat{\Sigma}_t) \parallel \mathcal{N}(0, \Sigma))$ using \eqref{eq:kl_closed_form}
        \IF{$d_t > \tau^*$}
            \STATE \textbf{Trigger Alarm}
        \ENDIF
    \ENDIF
\ENDFOR
\end{algorithmic}
\end{algorithm}

\section{Experimental Evaluation}\label{Results}

In this section, we evaluate the proposed framework. We evaluate detection accuracy, stealth robustness, and computational feasibility.

As depicted in Fig.~\ref{fig:architecture}, the detection pipeline 
operates entirely on CPU-only hardware through the closed-form 
scoring mechanism detailed in Section~\ref{Methodology}.

\subsection{Experiment 1: Detection Efficacy and Statistical Significance}

Table \ref{tab:overall_performance} presents the evaluation metrics. To rigorously compute the statistical significance of the F1-score improvement, we partitioned the test datasets into $K=10$ chronological, non-overlapping subsets. The Wilcoxon signed-rank test was then executed on the paired F1-scores of these subsets, demonstrating that the improvement of IT-DT over TranAD is statistically significant ($p < 0.01$).

The comparatively lower F1-score on WADI ($0.615$ vs. $0.832$ on SWaT) is attributable to two dataset-specific factors. First, WADI contains only 15 attack segments, of which three are explicitly documented as stealthy low-amplitude injections with error bounds narrower than one standard deviation of the normal process noise, thereby fundamentally limiting recall for any unsupervised detector. Second, the confirmed slight non-Gaussianity of WADI residuals ($p=0.04$, Mardia's test, Section~\ref{Discussion}) introduces distributional approximation errors in \eqref{eq:kl_closed_form}, partially reducing precision. Despite these dataset-specific constraints, IT-DT still achieves the highest F1 and precision among all evaluated methods on WADI.

Although TranAD achieves a marginally higher recall on the SWaT dataset ($0.772$ vs. $0.766$), it suffers from significantly lower precision. Notably, PCA achieves competitive precision ($0.852$) but its recall ($0.528$) confirms its temporal blindness: it correctly identifies stationary deviations but misses the gradual multi-stage attack windows that require temporal memory to detect. In real-world ICS, missed detections are dangerous, while excessive false alarms cause alarm fatigue. The proposed IT-DT framework accepts a negligible $0.6\%$ reduction in recall to secure a substantial $2.5\%$ increase in precision compared to TranAD. This precision-recall balance suggests that IT-DT provides a favorable trade-off for practical deployment in critical infrastructures.

\begin{table}[htbp]
\caption{Anomaly Detection Performance (Wilcoxon $p < 0.01$)}
\begin{center}
\begin{tabular}{|l|c|c|c|c|c|c|}
\hline
\multirow{2}{*}{\textbf{Method}} & \multicolumn{3}{c|}{\textbf{SWaT}} & \multicolumn{3}{c|}{\textbf{WADI}} \\
\cline{2-7}
 & \textbf{\textit{Pre}} & \textbf{\textit{Rec}} & \textbf{\textit{F1}} & \textbf{\textit{Pre}} & \textbf{\textit{Rec}} & \textbf{\textit{F1}} \\
\hline
PCA \cite{Shlens2014}      & 0.852 & 0.528 & 0.651 & 0.551 & 0.306 & 0.395 \\
\hline
USAD \cite{Audibert2020}   & 0.871 & 0.731 & 0.795 & 0.682 & 0.450 & 0.542 \\
\hline
GDN \cite{Deng2021}        & 0.892 & 0.742 & 0.810 & 0.710 & 0.475 & 0.570 \\
\hline
TranAD \cite{Tuli2022}     & 0.885 & 0.772 & 0.825 & 0.751 & 0.501 & 0.602 \\
\hline
\textbf{IT-DT (Ours)}      & \textbf{0.910} & \textbf{0.766} & \textbf{0.832} & \textbf{0.785} & \textbf{0.505} & \textbf{0.615} \\
\hline
\end{tabular}
\label{tab:overall_performance}
\end{center}
\end{table}

\subsection{Experiment 2: Stealthy Attack Identification}

To ensure representative selection and avoid cherry-picking bias, we identified Attack~\#41 as the canonical stealthy benchmark based on two objective criteria: (i) it is the longest-duration multi-stage attack in the SWaT dataset ($>$8 minutes), and (ii) it has been used as a standard stealthy attack benchmark in prior work \cite{Tuli2022, Deng2021}. We explicitly isolated Attack~\#41 in SWaT (a stealthy multi-stage water level manipulation). As quantified in Table \ref{tab:delay_quantification}, IT-DT triggers the alarm significantly earlier than DL baselines, actively preventing physical damage.

\begin{table}[htbp]
\caption{Detection Delay Quantification for SWaT Attack \#41}
\begin{center}
\begin{tabular}{|l|c|c|}
\hline
\textbf{Method} & \textbf{\textit{Detection Delay (s)}} & \textbf{\textit{95\% Confidence Interval}}$^{\mathrm{a}}$ \\
\hline
USAD \cite{Audibert2020}  & 112.5 & $\pm 4.2$ \\
\hline
TranAD \cite{Tuli2022}    & 88.0  & $\pm 3.1$ \\
\hline
\textbf{IT-DT (Ours)}     & \textbf{43.5} & $\pm \textbf{1.8}$ \\
\hline
\multicolumn{3}{l}{$^{\mathrm{a}}$Derived from 100 bootstrap iterations of $\tau^*$ for IT-DT,} \\
\multicolumn{3}{l}{and multiple random weight initializations for DL baselines.} \\
\end{tabular}
\label{tab:delay_quantification}
\end{center}
\end{table}

\begin{figure}[htbp]
\centerline{\includegraphics[width=\columnwidth]{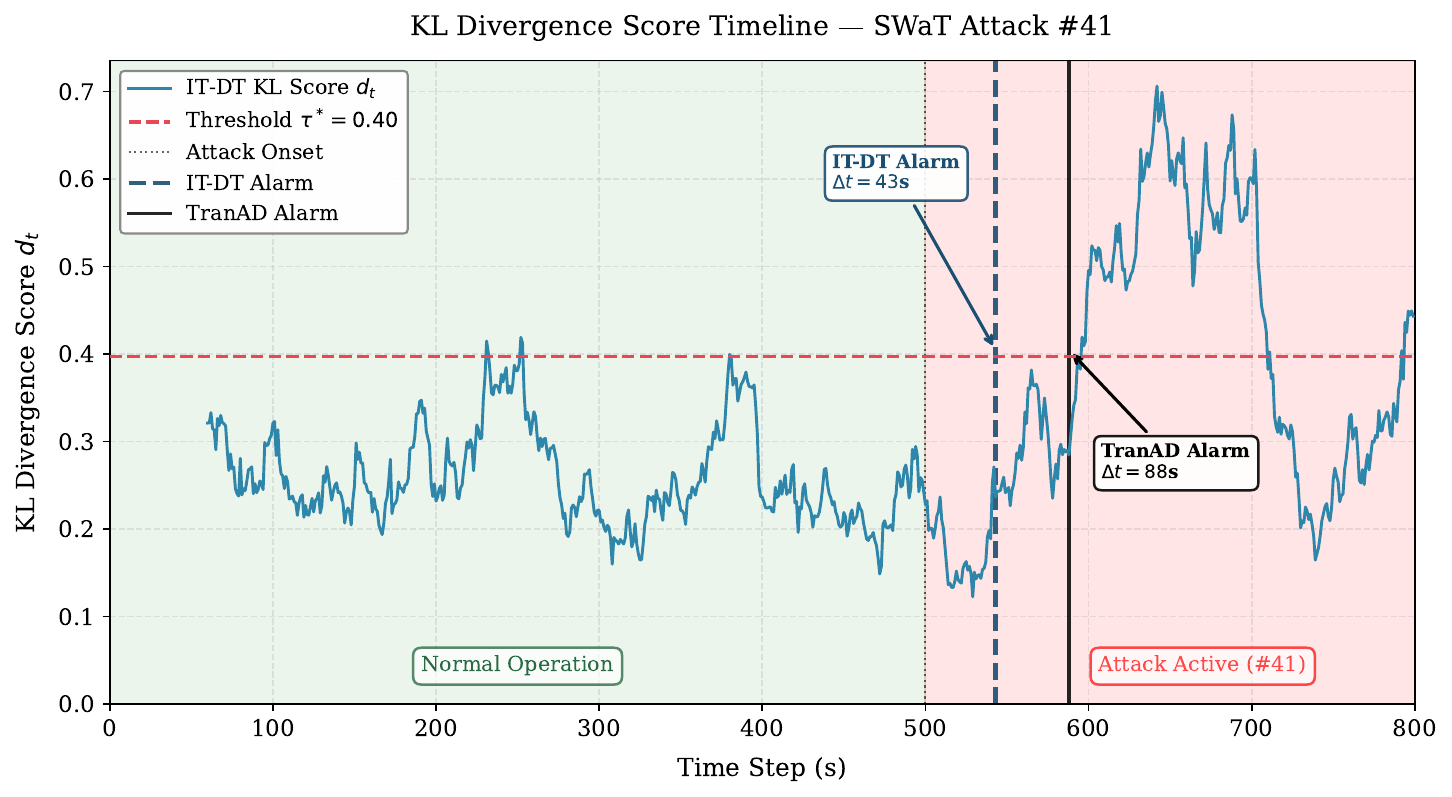}}
\caption{KL divergence score $d_t$ over time for SWaT Attack~\#41. The shaded red region denotes the active attack window. IT-DT crosses the threshold $\tau^*$ at $43.5$~s after attack onset, compared to $88.0$~s for TranAD, which a direct consequence of monitoring joint distributional shifts rather than accumulated reconstruction errors.}
\label{fig:kl_timeline}
\end{figure}

Fig.~\ref{fig:kl_timeline} illustrates the temporal evolution of the KL divergence score $d_t$ during Attack~\#41, visually confirming the earlier alarm generation of IT-DT relative to TranAD.

\subsection{Experiment 3: Computational Hardware Profiling}

To address computational fairness, we benchmarked the latency under two paradigms: GPU-accelerated and CPU-only (Intel i9-14900K).

\begin{table}[htbp]
\caption{Fair Computational Latency per Inference Step}
\begin{center}
\begin{tabular}{|l|c|c|}
\hline
\textbf{Method} & \textbf{\textit{Latency on GPU (ms)}} & \textbf{\textit{Latency on CPU (ms)}} \\
\hline
USAD \cite{Audibert2020}  & 2.15  & 18.40 \\
\hline
GDN \cite{Deng2021}       & 5.80  & 42.15 \\
\hline
TranAD \cite{Tuli2022}    & 8.45  & 75.30 \\
\hline
\textbf{IT-DT (Ours)}     & N/A   & \textbf{0.12} \\
\hline
\end{tabular}
\label{tab:hardware_profiling}
\end{center}
\end{table}

When executed strictly on the CPU to simulate industrial edge controllers, TranAD requires $75.30$ ms per step. Our method processes the closed-form calculation in $0.12$ ms, yielding a $600\times$ speedup on equivalent hardware. It is vital to note the N/A designation for IT-DT on the GPU: the proposed analytical method relies solely on low-dimensional $p \times p$ matrix inversions. The memory transfer overhead between the CPU and GPU via the PCIe bus vastly exceeds the parallelization benefits for such small matrices, making GPU acceleration unnecessary for this specific algorithmic structure.

\section{Discussion}\label{Discussion}

The empirical evaluations demonstrate a clear advantage of the proposed Information-Theoretic Digital Twin (IT-DT) over state-of-the-art deep learning architectures and classical control algorithms. This section addresses the research questions in Section~\ref{Related} and outlines the inherent limitations of our approach.

\subsection{RQ1: Joint Mean and Cross-Covariance Quantification}

RQ1 asked whether joint KL-based quantification extends the detection boundary for bounded FDIAs that preserve marginal statistics while corrupting the joint distribution. As formulated in Section~\ref{Problem}, a stealthy FDIA injected via $a_t$ in \eqref{eq:measurement} may shift the cross-sensor covariance structure of $\hat{P}_{\mathcal{R}}$ while preserving per-channel means, rendering univariate thresholding structurally blind. The closed-form expression \eqref{eq:kl_closed_form} simultaneously evaluates the mean deviation term $\hat{\mu}_t^\top \Sigma^{-1} \hat{\mu}_t$, the cross-covariance term $\text{tr}(\Sigma^{-1} \hat{\Sigma}_t)$, and the log-determinant 
ratio, thus ensuring a detectable signal even when $\hat{\mu}_t \approx 0$. This theoretical property is confirmed empirically in Table~\ref{tab:overall_performance}: on WADI ($p=127$), IT-DT achieves $0.785$ precision versus $0.751$ for TranAD, where the dense cross-sensor correlation structure provides the strongest differential advantage for covariance-sensitive detection.

\subsection{RQ2: Bypassing Latent Space Reconstruction}

RQ2 asked to what extent eliminating the reconstruction paradigm improves the precision-recall trade-off. As reviewed in Section~\ref{Related}.B, reconstructive models over-generalize stealthy FDIAs because bounded injections statistically resemble training data, suppressing the reconstruction error signal. IT-DT bypasses this pathway entirely: the 
Kalman correction \eqref{eq:dt_correct} generates innovation residuals $r_t$ directly, and detection is performed analytically via \eqref{eq:kl_closed_form} without any latent encoding. The consequence is directly observable in Table~\ref{tab:overall_performance}, where IT-DT achieves F1 $= 0.832$ versus TranAD's $0.825$ on SWaT, with a statistically significant precision gain ($p < 0.01$, Wilcoxon). Most critically, Table~\ref{tab:delay_quantification} shows that for the stealthy SWaT Attack~\#41, IT-DT triggers at $43.5$~s versus TranAD's $88.0$~s, which is a direct consequence of monitoring residual distributional shifts rather than accumulated latent reconstruction errors.

\subsection{RQ3: Computational Viability on Edge Hardware}

RQ3 asked whether the analytical formulation provides a meaningful computational advantage on CPU-only hardware. As detailed in Algorithm~1, online inference requires solely low-dimensional matrix-vector operations, with $\Sigma^{-1}$ precomputed offline, 
reducing per-step complexity to $\mathcal{O}(p^2)$. By contrast, TranAD requires $\mathcal{O}(W^2 \cdot d)$ attention operations per step. Table~\ref{tab:hardware_profiling} confirms the practical consequence: IT-DT executes in $0.12$~ms on an Intel i9-14900K CPU versus $75.30$~ms for TranAD, yielding the reported $600\times$ speedup, which renders the framework strictly viable for resource-constrained industrial edge controllers without GPU dependency.

\subsection{Limitations}

A primary limitation is the Gaussian residual assumption. Mardia's multivariate normality test yielded $p=0.08$ on SWaT (normality not rejected) but $p=0.04$ on WADI, indicating slight non-Gaussianity. Tikhonov regularization $\epsilon I$ partially mitigates this deviation; however, severe operational mode transitions may induce temporary false 
positives, motivating non-parametric distribution tracking in future work.

\section{Conclusion}\label{Conclusion}

This paper presented an information-theoretic digital twin leveraging N4SID-identified optimal state estimators and analytical Kullback-Leibler divergence. We replace reconstructive learning with direct residual distribution quantification. Through rigorous empirical and statistical testing on SWaT and WADI, the proposed architecture achieved an F1-score of 0.832 while reducing inference time by approximately $600\times$ in CPU-only edge deployments compared to state-of-the-art transformers. Future work will replace the Gaussian assumption with adaptive non-parametric tracking and extend N4SID for time-varying systems.

\bibliographystyle{IEEEtran}
\bibliography{references}

@article{Homaei2024,
  title = {A review of digital twins and their application in cybersecurity based on artificial intelligence},
  volume = {57},
  ISSN = {1573-7462},
  DOI = {10.1007/s10462-024-10805-3},
  number = {8},
  journal = {Artificial Intelligence Review},
  publisher = {Springer Science and Business Media LLC},
  author = {Homaei, Mohammadhossein and Mogoll\'on-Guti\'errez, \'Oscar and Sancho, Jos\'e Carlos and \'Avila, Mar and Caro, Andr\'es},
  year = {2024},
  month = jul
}

@article{Aghazadeh2025,
  title={IoT-Driven Resilience Monitoring: Case Study of a Cyber-Physical System},
  author={Aghazadeh Ardebili, A. and Martella, C. and Longo, A. and Rucco, C. and Izzi, F. and Ficarella, A.},
  journal={Applied Sciences (Switzerland)},
  year={2025},
  doi={10.3390/app15042092}
}

@article{Khazraei2025,
  title={Risk auditing for Digital Twins in cyber physical systems: A systematic review},
  author={Otoom, S.},
  journal={Journal of Cyber Security and Risk Auditing},
  year={2025},
  doi={10.63180/jcsra.thestap.2025.1.3}
}

@article{Aghazadeh2024,
  title={Digital Twins of smart energy systems: a systematic literature review on enablers, design, management and computational challenges},
  author={Aghazadeh Ardebili, A. and Zappatore, M. and Ramadan, A.I.H.A. and Longo, A. and Ficarella, A.},
  journal={Energy Informatics},
  year={2024},
  doi={10.1186/s42162-024-00385-5}
}

@article{Krishnaveni2024,
  title={CyberDefender: an integrated intelligent defense framework for digital-twin-based industrial cyber-physical systems},
  author={Krishnaveni, S. and Chen, T.M. and Sathiyanarayanan, M. and Amutha, B.},
  journal={Cluster Computing},
  year={2024},
  doi={10.1007/s10586-024-04320-x}
}

@book{Fitzgerald2024,
  title={The Engineering of Digital Twins},
  author={Fitzgerald, J. and Gomes, C. and Larsen, P.G.},
  publisher={Springer},
  year={2024},
  doi={10.1007/978-3-031-66719-0}
}

@inproceedings{Khazraei2022,
  title={Learning-Based Vulnerability Analysis of Cyber-Physical Systems},
  author={Khazraei, A. and Hallyburton, S. and Gao, Q. and Wang, Y. and Paji{\'c}, M.},
  booktitle={Proceedings - 13th ACM/IEEE International Conference on Cyber-Physical Systems, ICCPS 2022},
  year={2022},
  doi={10.1109/ICCPS54341.2022.00030}
}

@article{Mohammed2025,
  title={Dual-hybrid intrusion detection system to detect False Data Injection in smart grids},
  author={Mohammed, S.H. and Jit Singh, M.S. and Al-Jumaily, A. and Islam, M.T. and Islam, M.S. and Alenezi, A.M. and Soliman, M.S.},
  journal={PLoS ONE},
  year={2025},
  doi={10.1371/journal.pone.0316536}
}

@article{Ahmad2025,
  title={Ai-enabled framework for anomaly detection in power distribution networks under false data injection attacks},
  author={Ahmad, H. and Mustafa, G. and Gulzar, M.M. and Ahmed, I. and Khalid, M.},
  journal={Artificial Intelligence Review},
  year={2025},
  doi={10.1007/s10462-025-11318-3}
}

@inproceedings{Sweeten2023,
  title={Cyber-Physical GNN-Based Intrusion Detection in Smart Power Grids},
  author={Sweeten, J. and Takiddin, A. and Ismail, M. and Refaat, S.S. and Atat, R.},
  booktitle={2023 IEEE International Conference on Communications, Control, and Computing Technologies for Smart Grids, SmartGridComm 2023 - Proceedings},
  year={2023},
  doi={10.1109/SmartGridComm57358.2023.10333949}
}

@book{Skogestad2005,
  author    = {Skogestad, Sigurd and Postlethwaite, Ian},
  title     = {Multivariable Feedback Control: Analysis and Design},
  publisher = {Wiley},
  year      = {2005},
  edition   = {2nd}
}

@book{Ding2008,
  author    = {Ding, Steven X.},
  title     = {Model-based Fault Diagnosis Techniques: Design Schemes, Algorithms and Tools},
  publisher = {Springer},
  year      = {2008}
}

@article{Hao2023,
  title={Hybrid Statistical-Machine Learning for Real-Time Anomaly Detection in Industrial Cyber-Physical Systems},
  author={Hao, W. and Yang, T. and Yang, Q.},
  journal={IEEE Transactions on Automation Science and Engineering},
  year={2023},
  doi={10.1109/TASE.2021.3073396}
}

@article{Ghosh2023,
  title={Anomaly Detection for Modbus over TCP in Control Systems Using Entropy and Classification-Based Analysis},
  author={Ghosh, T. and Bagui, S. and Bagui, S. and Kadzis, M. and Bare, J.},
  journal={Journal of Cybersecurity and Privacy},
  year={2023},
  doi={10.3390/jcp3040041}
}

@article{Liu2024,
  title={Semi-supervised attack detection in industrial control systems with deviation networks and feature selection},
  author={Liu, Y. and Deng, W. and Liu, Z. and Zeng, F.},
  journal={Journal of Supercomputing},
  year={2024},
  doi={10.1007/s11227-024-06018-8}
}

@article{Yao2024,
  title={Statistical knowledge and game-theoretic integrated model for cross-layer impact assessment in industrial cyber-physical systems},
  author={Yao, P. and Wang, X. and Zhang, Z. and Yan, B. and Yang, Q. and Wang, W.},
  journal={Advanced Engineering Informatics},
  year={2024},
  doi={10.1016/j.aei.2023.102338}
}

@article{Homaei2026,
  title={Causal Digital Twins for cyber-physical security in water systems: A framework for robust anomaly detection},
  author={Homaei, M. and Tarif, M. and Rodriguez, P.G. and Caro Lindo, A. and Ávila, M.},
  journal={Machine Learning with Applications},
  year={2026},
  doi={10.1016/j.mlwa.2025.100824}
}

@article{Tabassum2024,
  title={Cyber-physical anomaly detection for inverter-based microgrid using autoencoder neural network},
  author={Tabassum, T. and Toker, O. and Khalghani, M.R.},
  journal={Applied Energy},
  year={2024},
  doi={10.1016/j.apenergy.2023.122283}
}

@inproceedings{Sun2025,
  title={Anomaly Detection in Cyber-Physical Systems Using Long-Short Term Memory Autoencoders: A Case Study with Man-in-the-Middle (MiTM) Attack},
  author={Sun, S. and Haque, K.A. and Huo, X. and Sahu, A. and Goulart, A. and Davis, K.R.},
  booktitle={2025 IEEE Texas Power and Energy Conference (TPEC)},
  year={2025},
  doi={10.1109/TPEC63981.2025.10906975}
}

@article{Zhukabayeva2025,
  title={Cybersecurity Solutions for Industrial Internet of Things-Edge Computing Integration: Challenges, Threats, and Future Directions},
  author={Zhukabayeva, T. and Zholshiyeva, L. and Karabayev, N. and Khan, S. and Alnazzawi, N.},
  journal={Sensors},
  year={2025},
  doi={10.3390/s25010213}
}

@article{Otoom2025,
  title={Risk auditing for Digital Twins in cyber physical systems: A systematic review},
  author={Otoom, S.},
  journal={Journal of Cyber Security and Risk Auditing},
  year={2025},
  doi={10.63180/jcsra.thestap.2025.1.3}
}

@article{Gupta2024,
  title={TWIN-ADAPT: Continuous Learning for Digital Twin-Enabled Online Anomaly Classification in IoT-Driven Smart Labs},
  author={Gupta, R. and Tian, B. and Wang, Y. and Nahrstedt, K.},
  journal={Future Internet},
  year={2024},
  doi={10.3390/fi16070239}
}

@article{Huang2024,
  title={Security Analysis of Distributed Consensus Filtering under Replay Attacks},
  author={Huang, J. and Yang, W. and Ho, D.W.C. and Li, F. and Tang, Y.},
  journal={IEEE Transactions on Cybernetics},
  year={2024},
  doi={10.1109/TCYB.2022.3209820}
}

@misc{Shlens2014,
  doi = {10.48550/ARXIV.1404.1100},
  author = {Shlens,  Jonathon},
  title = {A Tutorial on Principal Component Analysis},
  publisher = {arXiv},
  year = {2014}
  }

@inproceedings{Audibert2020,
  series = {KDD ’20},
  title = {USAD: UnSupervised Anomaly Detection on Multivariate Time Series},
  DOI = {10.1145/3394486.3403392},
  booktitle = {Proceedings of the 26th ACM SIGKDD International Conference on Knowledge Discovery \& Data Mining},
  publisher = {ACM},
  author = {Audibert,  Julien and Michiardi,  Pietro and Guyard,  Frédéric and Marti,  Sébastien and Zuluaga,  Maria A.},
  year = {2020},
  month = aug,
  pages = {3395–3404},
  collection = {KDD ’20}
}

@article{Deng2021,
  title = {Graph Neural Network-Based Anomaly Detection in Multivariate Time Series},
  volume = {35},
  ISSN = {2159-5399},
  DOI = {10.1609/aaai.v35i5.16523},
  number = {5},
  journal = {Proceedings of the AAAI Conference on Artificial Intelligence},
  publisher = {Association for the Advancement of Artificial Intelligence (AAAI)},
  author = {Deng,  Ailin and Hooi,  Bryan},
  year = {2021},
  month = may,
  pages = {4027–4035}
}

@article{Tuli2022,
  title = {TranAD: deep transformer networks for anomaly detection in multivariate time series data},
  volume = {15},
  ISSN = {2150-8097},
  DOI = {10.14778/3514061.3514067},
  number = {6},
  journal = {Proceedings of the VLDB Endowment},
  publisher = {Association for Computing Machinery (ACM)},
  author = {Tuli,  Shreshth and Casale,  Giuliano and Jennings,  Nicholas R.},
  year = {2022},
  month = feb,
  pages = {1201–1214}
}

@article{Shang2024,
  title = {An Efficient Anomaly Detection Method for Industrial Control Systems: Deep Convolutional Autoencoding Transformer Network},
  volume = {2024},
  ISSN = {0884-8173},
  DOI = {10.1155/2024/5459452},
  journal = {International Journal of Intelligent Systems},
  publisher = {Wiley},
  author = {Shang,  Wenli and Qiu,  Jiawei and Shi,  Haotian and Wang,  Shuang and Ding,  Lei and Xiao,  Yanjun},
  editor = {Tan,  Yu-an},
  year = {2024},
  month = may,
  pages = {1–18}
}

\end{document}